\documentclass{aa}
\RequirePackage{etex}
\usepackage{color} 
\usepackage{graphicx}
\usepackage{txfonts}
\usepackage{lipsum}
\usepackage{subcaption}         % necessary for continued figures, example in section 3
\usepackage{lscape}             % to rotate a single page table, example in appendix.
\usepackage{placeins}           % useful with \FloatBarrier, to keep 
\begin{document}

   \title{Radio Detection of a Local Little Red Dot}

%   \subtitle{Subtitle}

%%%%%%%%%%%%%%%%%%%%%%%%%%%%%%%%%%%%%%%%
% Please do not include ORCIDs next to author names.
% Only ORCIDs authenticated by individual authors in EDP Sciences editorial system will be taken into account.
% ORCIDs included here will be removed.
%%%%%%%%%%%%%%%%%%%%%%%%%%%%%%%%%%%%%%%%

   \author{L. F. Rodriguez\inst{1,2}
        \and I. F. Mirabel\inst{3,4}
%        \fnmsep\thanks{Shows the usage of elements in the author field}
        }
   \institute{Instituto de Radioastronomía y Astrofísica, Universidad Nacional Autónoma de México, Apdo. Postal 3-72, Morelia, Michoacán 58089, México \\
             \email{l.rodriguez@irya.unam.mx}
%             \thanks{Shows the usage of elements in the author field}
            \and Mesoamerican Center for Theoretical Physics, Universidad Autónoma de Chiapas, Tuxtla Gutiérrez, Chiapas 29050, México
            \and Département d’Astrophysique-IRFU-CEA Université Paris-Saclay, France
            \and Instituto de Astronomía y Física del Espacio (IAFE) CONICET-Universidad de Buenos Aires, C1428 Buenos Aires, Argentina \\}

   \date{Received November 29, 2025}

% \abstract{}{}{}{}{}
% 5 {} token are mandatory
 
  \abstract
  % context heading (optional)
  % {} leave it empty if necessary  
   {One of the most important discoveries by the James Webb Space Telescope (JWST) is the unexpected existence in the Early Universe (z > 4) of very large quantities of “Little Red Dots” (LRDs), compact luminous red galaxies of intriguing physical properties. 
     One of those intriguing properties is the absence of radio detections in high redshift LRDs.    
}
  % aims heading (mandatory)
   { 
   We wish to know if LRDs have radio emission that may be produced by accreting Intermediate/Supermassive Black Holes (IMBHs/SMBHs) or by frequent supernovae (SNe) from 
   a cluster of massive stars. 
    }
%   The spectrum of radio emission (synchrotron vs thermal) can be used to know % % which between these two types of energy sources is the dominant one.} 
  % methods heading (mandatory)
   {  Assuming LRDs at high redshifts have not been detected at radio wavelengths due to their large distances and/or present limitations of observational capabilities, we analyse here
   archive Very Large Array radio observations  
   of J1047+0739 and J1025+1402,  
   two analog candidates of LRDs in the Local Universe (LLRDs) at redshifts z = 0.1 - 0.2. }
  % results heading (mandatory)
   {The LLRD source J1047+0739 at z = 0.1682 is detected at 6.0 GHz in 2018 with
    the VLA (Very Large Array)
    as a compact source with radius less than 0.2 arcsec ($<$700 pc at d $\simeq$ 750 Mpc). Its flux density was 117$\pm$8 $\mu$Jy and its 
     in-band spectral index 
   was --0.85$\pm$0.24,   which is typical of 
   optically-thin synchrotron emission. It was also detected at 5.0 GHz in 2010
   with the VLA-C, showing a flux density of 130$\pm$9 $\mu$Jy.}
  % conclusions heading (optional), leave it empty if necessary
   {The observed flux densities can be provided by either a radio luminous supernova
   or an accreting IMBH/SMBH. However, the lack of important variation 
   in flux density over eight years favors the IMBH/SMBH hypothesis. Radio
   time monitoring of this and other LLRDs could help clarify the mystery of the
   radio silence of its cosmological counterparts.}

\maketitle

%%%%%%%%%%%%%%%%%%%%%%%%%%%%%%%%%%%%%%%%%%%%%%%%%%%%%%%%%%%%%%
\section{Introduction}
%\lipsum[1]
%%%%%%%%%%%%%%%%%%%%%%%%%%%%%%%%%%%%%%%%%%%%%%%%%%%%%%%%%%%%%%
Abundant populations of LRDs at z > 4 have been unexpectedly
discovered in several JWST surveys in the last two years (e.g., Matthee et al. 2024;   Kocevski et al. 2025a; Labbe et al. 2025; Akins et al. 2025).   LRDs are mostly observed between redshifts of z = 4 and z = 8 with maximum peak density at z = 5, and an exponential decline in numbers at redshifts z < 4 (e.g., Billand et al. 2025), which implies that Local Little Red Dot (LLRD) candidates are scarce in the Local Universe. 

LRDs in the Early Universe are extremely compact galaxies with average effective radius of less than a few hundred pc, widths of broad emission lines of up to 2000 km s$^{-1}$, and V-shaped Spectral Energy Distributions, intrinsically red rather than dust-reddened. These 
properties together are uncommon in previously known galaxies before JWST. 
LRDs are not detected or are very dim soft X-ray emitters of less than 10 keV, and have not been detected at radio wavelengths. These physical properties are puzzling, and to explain the dominant energy source of the high luminosities there were proposed two basic kinds of models; one based on highly accreting IMBH/SMBHs, and another based on 
rich clusters of very massive stars. These latter models have been challenged 
because of the extreme stellar masses that would be required to produce the high luminosity of LRDs in volumes with radii of only a few hundred pc (Pacucci et al. 2023).
On the other hand, the absence of soft X-ray detections in LRDs challenged the accreting IMBH/SMBH models. 
However, this absence of X-rays can be explained by cold gas absorption in an IMBH/SMBH 
atmosphere with gas densities $\rm n_{\rm H} > 10^9 ~cm^{-3}$, column densities $\rm N_{\rm H} > 10^{24}~ cm^{-2}$, and gas turbulent velocities of $\sim$500 km s$^{-1}$ (e.g., Maiolino et al. 2025;
 
Kocevski et al. 2025b). 
Other models attempted to explain the absence of X-rays and radio emission by dust or   
free-free absorption, synchrotron self-absorption, and disruption of magnetic field and X-ray corona by super-Eddington accretion (Mazzolari et al. 2024;   de Graaff et al. 2025).   

Contrary to soft X-rays, radio jet emission may not be totally or even partially absorbed by 
neutral column densities of  $\rm N_{\rm H} > 10^{24}~ cm^{-2}$. In fact, radio jets are observed from mass accreting stellar BHs while the soft X-rays are completely obscured (Rodriguez and Mirabel 2025 and references therein). In addition, we have shown that under
the assumption that radio jets are perpendicular to the inner BH accretion disks, radio jets can provide crucial information on the physics of accretion in the
inner obscured disks. 
However, LRDs in the Early Universe so far have been radio silent due to unknown reasons (e.g., Mazzolari et al. 2024; Perger et al. 2025; Akins et al. 2025; Latif et al. 2025). In this context, the study of LLRDs analogs may help to understand why LRDs at high redshifts are radio silent, and eventually lead in the future to strategies that
may result in radio detections of LRDs in the Early Universe. 

An immediate possibility to test directly between the different hypotheses is the analysis of archived radio observations of LLRD candidates. 
 
The sources SDSS J1025+1402 (z = 0.10067)
and SDSS J1047+0739 (z = 0.16828) were serendipitously identified
for the first time in the Sloan Digital Sky Survey (SDSS) by Izotov \& Thuan (2008) as “Metal Poor Dwarf Emission-Line Galaxies”, finding in these galaxies signals of an intermediate-mass AGN (Active Galactic Nucleus).
The parameters of these sources are given in Table 1 of Izotov \& Thuan (2008).
It was found that these compact dwarf galaxies had extraordinarily velocity broadened 
H$\alpha$ emission with luminosities ranging from $3 \times 10^{41}$ to $2 \times 10^{42}$ erg s$^{-1}$. Their metallicity is very low and the extraordinarily high broad H$\alpha$ luminosities remained constant over periods of 3-7 yr, which probably excluded Type IIn supernovae as a mechanism for the emission with broad velocities. 
Further studies on this type of galaxies were made by Simmonds et al. (2016).

More recently, Lin et al. (2025) showed that these two sources exhibit properties fully consistent with those of high-redshift LRDs and
could be classified as local analogues of the LRDs. From new Chandra observations they found that these LRDs are very weak in X-rays, and quote several attempts of radio observations that resulted in no detections.

In section 2 the observations and data reduction are described.
In section 3 a discussion of the results is presented and in section 4
we summarize our results and present our conclusions.
%_____________________________________________________________
%                     Onecolumn continued float (place early!)
%-------------------------------------------------------------
\section{Observations and Data Reduction}

Previous studies of these regions failed to detect
radio continuum emission (Burke et al. 2021).
  
We searched for observations of these two LLRDs in the NRAO archive\footnote{The National 
Radio Astronomy Observatory is a facility of the National Science Foundation operated
under cooperative agreement by Associated Universities, Inc.}. We found data in two
VLA projects: 10B-156 (PI: C. Henkel) and 18A-413 (PI: F. Bauer). Both projects were observed in the VLA standard continuum mode (NRAO 2025).
 
The data were analyzed in the standard manner using the 
CASA (Common Astronomy Software Applications;  McMullin et al. 2007) package of NRAO and
the pipeline provided for VLA observations. We made images using a robust weighting (Briggs 1995) of 0 to optimize the compromise between angular resolution and sensitivity.
All images were also corrected for the primary beam
response.

In project
10B-156 both sources were observed. The array was in its C
configuration. J1047+0739 was observed at epoch 2010 Nov 06
and J1025+1402 was observed at epoch 2010 Oct 25. In both epochs
0521+166=3C138 was the amplitude calibrator. The gain calibrators
were J1007+1356 for J1025+1402 and J1058+0133 for J1047+0739.
The bootstrapped flux densities of 
J1007+1356 and J1058+0133 were 0.763$\pm$0.002 and  3.010$\pm$0.003 Jy,
respectively. Each target was observed for an on-source integration time of 88 minutes.

The data of project 18A-413 were taken with the array in its highest
angular resolution A configuration. Only J1047+0739 was observed, on epoch
2018 Apr 20.
The amplitude calibrator was 1331+305=3C286 and the gain calibrator was
J1058+0133, with a bootstrapped flux density of 6.379$\pm$0.024 Jy.

\subsection{LLRD J1047+0739}

The LLRD J1047+0739 was clearly 
detected in the first epoch (2010 Nov 06)
(Figure 1; top). After solving a problem with our initial amplitude calibration of the data for the project 10B-156, we infer for this source in epoch 2010 Nov 06, a flux density of 130$\pm$9 
$\mu$Jy.
Other parameters of the observations of this source are listed in Table 1.
The position, flux density and angular dimensions of the sources were
determined using the task \bf imfit \rm of CASA.  
In the second epoch (2018 Apr 20) J1047+0739
was observed for an on-source integration time of 19 minutes. Again, the source was
clearly detected (Figure 1; bottom), with a flux density of 117$\pm$8 $\mu$Jy. These
data were taken with the broad band correlator (with a total bandwidth of 4.0 GHz)
and an in-band spectral index determination was possible. For this we divided the
spectral windows in four sub-bands 1 GHz wide each and determined the
flux densities shown in Figure 2. A least-squares fit to these flux densities gives
$$\rm {log[S_\nu(\mu Jy)] = (2.74\pm0.18) -(0.85\pm0.24)~log[\nu(GHz)]}.
  \eqno{(1)}  $$
  
\subsection{LLRD J1025+1402}

 On the
other hand, no source was detected directly associated with J1025+1402
in the 2010 Oct 25 observations made as part of project 10B-156. However, about 2
arcsec to its east there is a source detected at a 4-$\sigma$ level with
a flux density of 45$\pm$10 $\mu$Jy (Figure 3). Following the formulation of
Anglada et al. (1998) the \sl a priori \rm probability of detecting a source with this
flux density at the frequency of 5.0 GHz in a solid angle of
4$\times$4 arcsec$^2$ is only 0.0005, so the association of the radio source with J1025+1402 seems
to be significant. 
We adopt a 4-$\sigma$ detection threshold for our targeted search given that the solid angle considered is small. A 4-$\sigma$ threshold has a false probability of 
$6.4 \times 10^{-5}$ per resolution element. Assuming a 
Nyquist sampling, the solid angle considered has 
9 resolution elements. The  probability of a false detection is then only 0.0006.

The radio source is unresolved at
a scale of $0\rlap.{''}9$ ($<$2000 pc at d $\simeq$ 450 Mpc).
 
This and other cosmological parameters were obtained using the
Javascript calculator from Wright (2006) and
assuming a flat Planck $\Lambda$CDM cosmology with
$\rm H_0 = 67.4\,\mathrm{km\,s^{-1}\,Mpc^{-1}}$,
$\Omega_M = 0.315$, and
$\Omega_\Lambda = 0.685$ (Planck Collaboration et al. 2020).

Future observations in VLA-A and VLA-B configurations may answer whether this extended radio emission may be a kpc jet emanating from an IMBH/SMBH.

%, and will pursue the study of this source in the future.

%_____________________________________________________________
%                        A figure as large as the column width
%-------------------------------------------------------------
   \begin{figure}[h!]
   \centering
   \vskip-1.0cm
   \hskip-1.6cm
   \includegraphics[width=1.15\hsize]{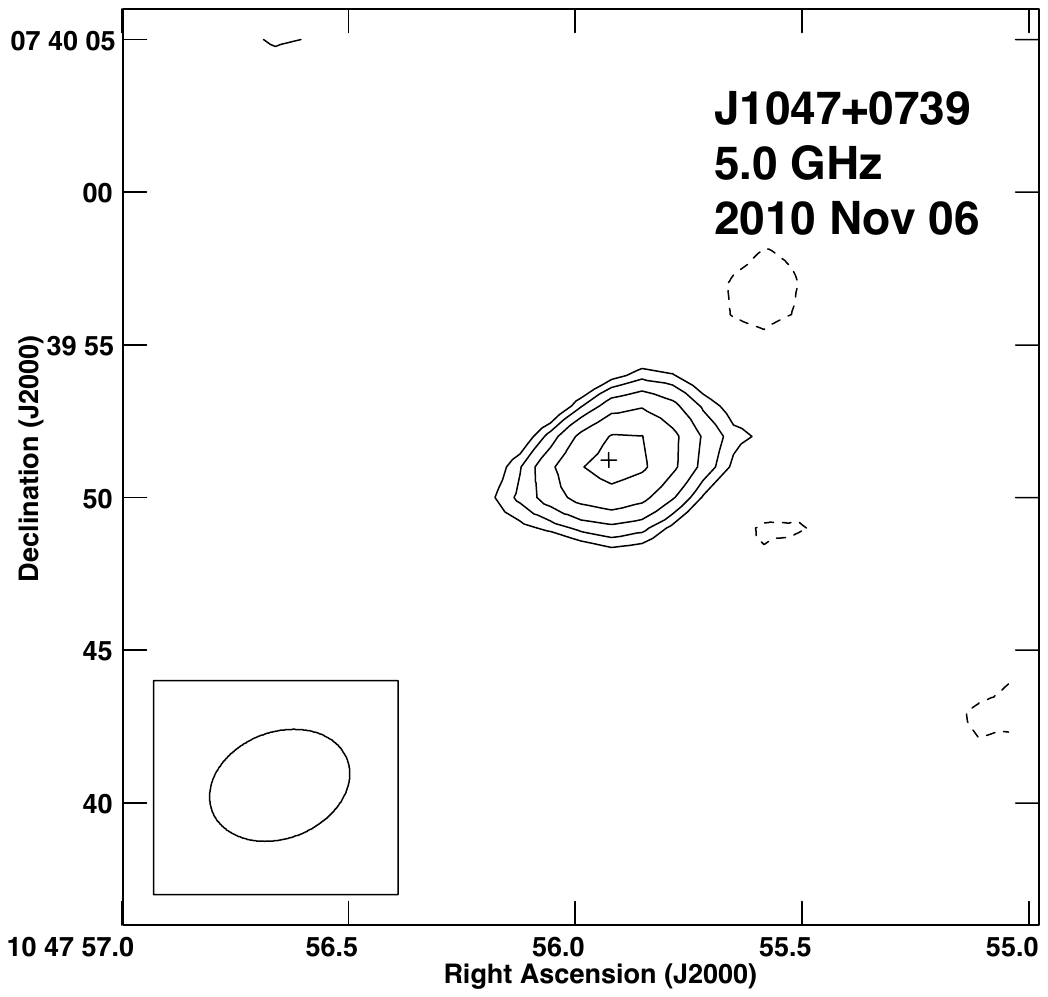}
  \vskip-5.5cm
  \hskip-1.8cm
   \includegraphics[width=1.2\hsize]{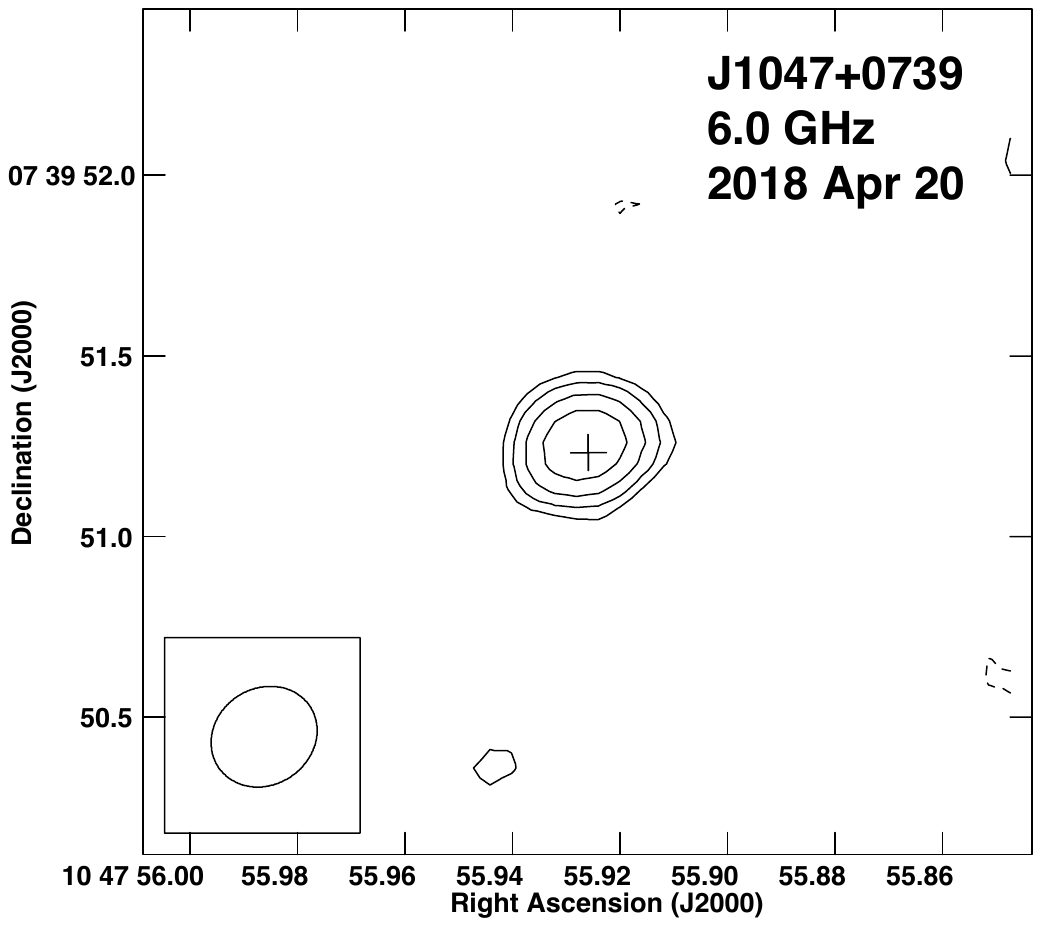}
 \vskip-2.8cm
      \caption{(Top) VLA contour image of the J1047+0739 region from project 10B-156.
  Contours start 
at $\pm$3-$\sigma$ and increase by factors of $\sqrt{2}$, 
      where $\sigma$ = 9.0 $\mu$Jy beam$^{-1}$,  
      the rms in this region of the image.
      The synthesized beam 
      ($4\rlap.^{''}7 \times 3\rlap.^{''}5; -69^\circ$)
      is shown in the bottom left corner.
      (Bottom) VLA contour image of the J1047+0739 region from project 18A-413.
        Contours start at $\pm$3-$\sigma$ and 
      increase by factors of $\sqrt{2}$, 
      where $\sigma$ = 8.0 $\mu$Jy beam$^{-1}$,  
      the rms in this region of the image.
      The synthesized beam ($0\rlap.^{''}30 \times 0\rlap.^{''}27; -56^\circ$)
      is shown in the bottom left corner.
      In both images the cross marks the optical position of the source from the
      Gaia Early Data Release 3 (Gaia Collaboration 2020).}          
         \label{fig1}
   \end{figure}
%
%_____________________________________________________________
%                     Long table using the longtab environment
%-------------------------------------------------------------
\longtab[1]{
\scriptsize
\begin{longtable}{ccccccccc}
\caption{VLA Observations of J1047+0739}\\
\label{longtable1}\\
\hline\hline
 &  & Frequency & Bandwidth & Spectral & & & & Flux Density \\
Project & Epoch & (GHz) & (GHz) & Windows & Channels & RA(J2000) & DEC(J2000) & ($\mu$Jy) \\
\hline
%\endfirsthead
%\caption{continued.}\\
%\hline\hline
% & & Frequency & & & \\
%Project & Epoch & (GHz) & Distance & Mode & Count Rate \\
%\hline
%\endhead

%\endfoot
%%
10B-156 & 2010 Nov 06 & 5.0 & 0.256 & 2 & 64 & 10 47 55.899 +/- 0.016 
& 07 39 51.20 +/- 0.12 & 130 +/- 9 \\
18A-413 & 2018 Apr 20 & 6.0 & 4.0 & 32 & 64 & 10 47 55.927 +/- 0.001 
& 07 39 51.26 +/- 0.01 & 117 +/- 8 \\
\hline
\end{longtable}
}

%%-------------------------------------------------------------
   \begin{figure}[h!]
   \centering
   \hskip-1.2cm
   \includegraphics[width=1.1\hsize]{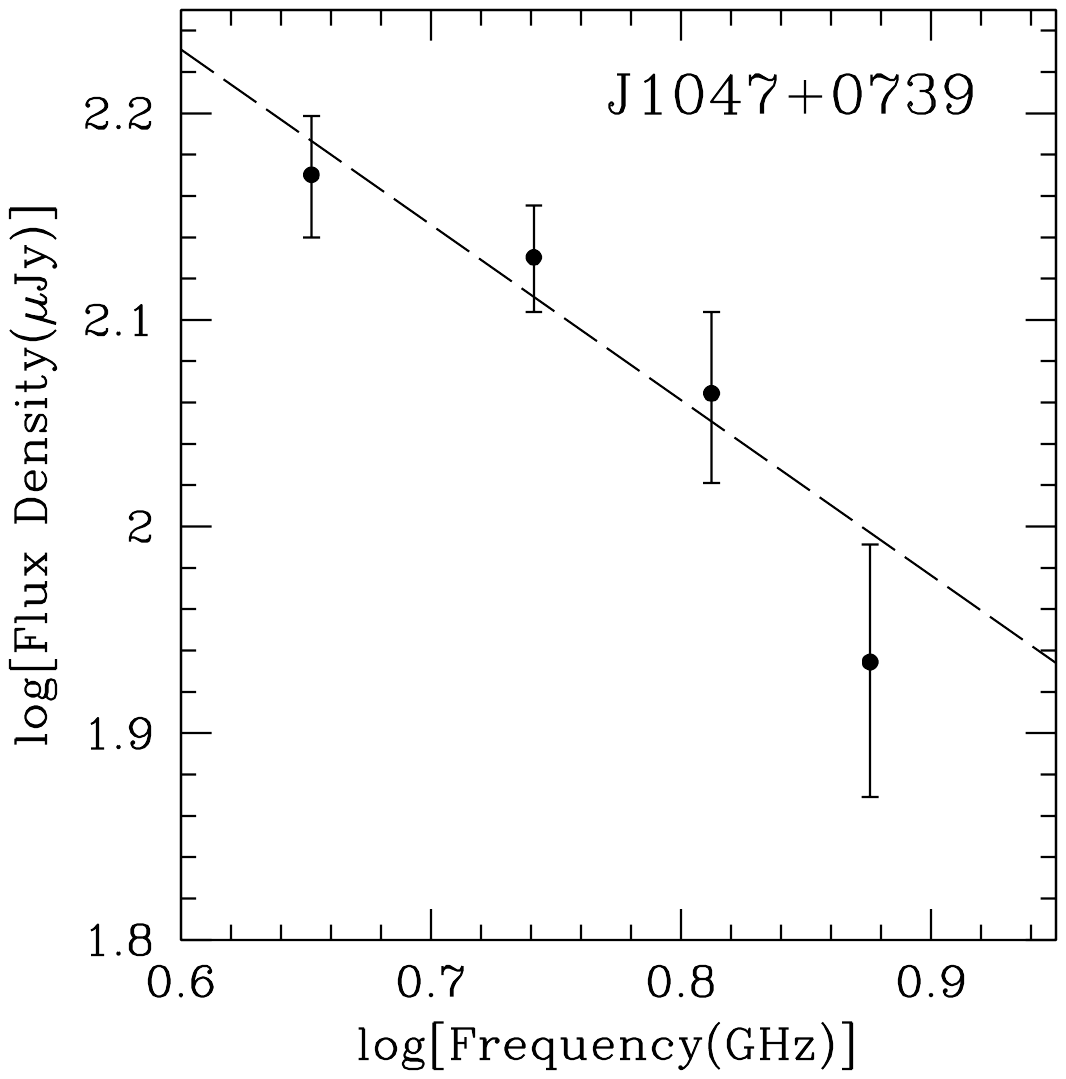}
   \vskip-2.5cm
      \caption{Flux density as a function of frequency for J1047+0739 from the data of project 18A-413.
       The least squares fit is indicated with a dashed line.}
%      gives 
%      log[{\rm S$_\nu$($\mu$Jy)] = (2.74$\pm$0.18) -(0.85$\pm$0.24) log[$%% %%%\nu$(GHz)}]. 
%      The spectral index of --0.85 is typical of optically-thin synchrotron 
%      emission.

         \label{fig2}
   \end{figure}
%
%-------------------------------------------------------------
%-------------------------------------------------------------
   \begin{figure}[h!]
   \centering
   \vskip-2.2cm
   \hskip-1.8cm
   \includegraphics[width=1.2\hsize]{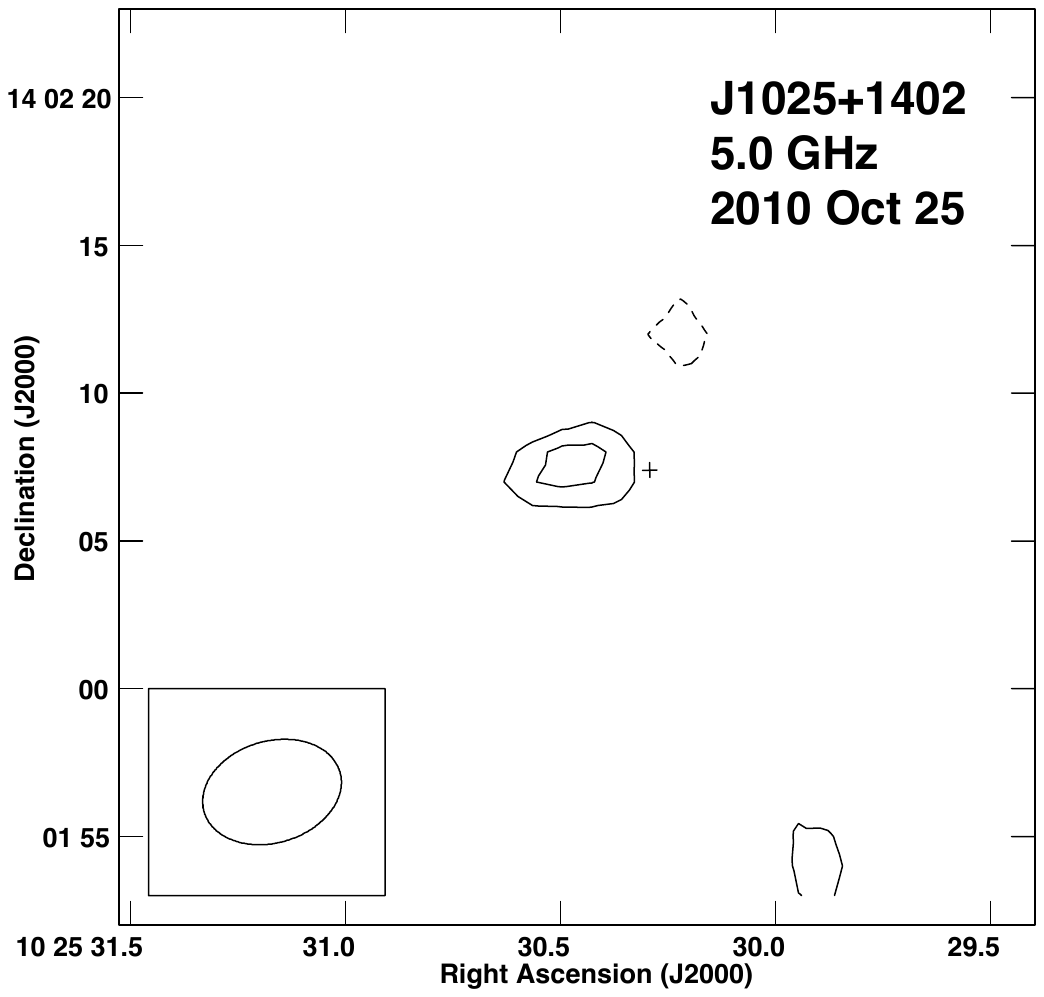}
   \vskip-2.5cm
      \caption{VLA contour image of the J1025+1402 region from project 10B-156.
        Contours start at $\pm$3-$\sigma$ and 
      increase by factors of $\sqrt{2}$, 
      where $\sigma$ = 10.0 $\mu$Jy beam$^{-1}$,  
      the rms in this region of the image. 
      The synthesized beam ($4\rlap.^{''}8 \times 3\rlap.^{''}4; -73^\circ$)
      is shown in the bottom left corner.
      The cross marks the optical position of the source from the
      Gaia Early Data Release 3 (Gaia Collaboration 2020).}
         \label{fig2}
   \end{figure}

\section{Discussion}

\subsection{The Nature of the Radio Emission of LLRD J1047+0739}

The spectral index of --0.85 found in J1047+0739
is typical of optically-thin synchrotron emission
and favors a non-thermal mechanism to produce the radio emission and possibly energize the
LRDs. The radio source is unresolved at
a scale of $0\rlap.{''}2$ ($<$700 pc at d $\simeq$ 750 Mpc).

Another important result is the lack of significant flux density variability in the
radio emission. Correcting the 5.0 GHz flux density of 2010 by a factor
$(5.0/6.0)^{-0.85}$ to compare with the 2018 flux density at 6.0 GHz,
we find that the flux density has remained approximately constant.
The ratio of flux densities 2018/2010 is 0.8$\pm$0.1 over this 
7.5 yr time period. 
%A detectable
%variability over a timescale of years is not uncommon in
%AGNs and again points to a SMBH, but as we will see, another explanation 
%is favored. 
This lack of variability in radio flux density is consistent with the lack
of variability at r-band reported for J1047+0739 at the 3-4\% level by
Burke et al. (2025)   over a rest-frame baseline of $\sim$15 yrs.

\subsection{Can a Supernova Account for the Radio Emission Observed in J1047+0739?}

%Can the radio emission detected in LLRD J1047+0739 be explained by a supernova?
The most radio luminous supernovae reach values of $L_R \sim 10^{39}$ erg s$^{-1}$
(Kulkarni et al. 1998; Yao et al. 2022). The radio luminosity associated with 
LLRD J1047+0739 between the rest frequencies $\nu_2$ and $\nu_1$ is calculated using
the following equation:

$$\rm {L_R = 4 \pi ~ D_L^2 ~ S_\nu ~ (1+z)^{-(\alpha+1)} \nu^{-\alpha} (\nu_2^{\alpha + 1}} - 
\nu_1^{\alpha + 1})/(\alpha + 1). \eqno{(2)}$$

\noindent The radio luminosity has been estimated by assuming that the 
power-law spectrum, $S_\nu \propto \nu^{\alpha}$, 
is valid between 1 and 10 GHz, that is that
$\nu_2$ = 10 GHz and $\nu_1$ = 1 GHz. 
In this frequency 
range optically-thin synchrotron emission dominates and spectral curvature 
is minimal (e.g. Condon 1992).
We have
an observed flux density of $\rm {S_\nu}$ = 117 $\mu$Jy at $\nu$ = 6.0 GHz.
For $\rm z$ = 0.16828, $\rm {D_L}$ = 839 Mpc and $\alpha$ = --0.85
we obtain $\rm{S_L} = 1.5 \times 10^{39}$ erg s$^{-1}$, a value comparable to that
of the most luminous radio supernovae known. We then conclude that the radio emission
observed from LLRD J1047+0739 can in principle be explained in terms of a luminous supernova.

However, the lack of significant variability between 2010 and 2018 argues against
an individual supernova mechanism.
%This could be explained in terms of two independent supernovae
%remnants. In this case the
%supernova rate in J1047+0739 would be enormous. However, a single supernova could
%also account for the observations. It is well known that supernova can suffer radio
%re-brightening as the ejecta interacts with surrounding gas even years or decades
%after the explosion (Balasubramanian et al. 2021; Rose et al. 2024; Soria et al. 2025).
%Monitoring of the radio emission will help understand if J1047+0739 has a high
%rate of supernovae or if we are observing a single source. 
%As a
%comparison, the starburst galaxy M82 is estimated to have a supernova rate of $\sim$0.1 yr$^{-1}$ %(one every ten years) occurring 
%over a region with $\sim$500 pc in diameter (Strickland 2004; Bolatto et al. 2024).
The radio emission could be produced by the combined effect of generations of supernova
explosions that injected the medium with relativistic electrons. 

\subsection{Can an IMBH/SMBH Account for the Radio Emission Observed in J1047+0739?}

The radio luminosity of
Seyfert galaxies is similar to that of J1047+0739. The radio luminosity functions of Seyfert galaxies of both types are similar. From a sample of 47 galaxies (Rush et al. 1996)
we find that its logarithmic mean and dispersion are given by $\rm log(L_R) = 10^{38.2 \pm 0.8}~erg~s^{-1}$. The radio luminosity of J1047+0739 falls in the high end of this range of values.
We conclude that an IMBH/SMBH of moderate activity can explain the observed radio luminosity of 
J1047+0739.

\subsection{Expected Radio Flux Density at Cosmological Distances}

We can use the measured flux density of LLRD J1047+0739 to predict the flux density
expected for an identical source located at cosmological distances.
For a radio source with a power-law spectrum given by 
$\rm{S_\nu \propto \nu^\alpha}$
the flux density measured at the same frequency and different $\rm z$ values is given by:

$$\Biggl[{{S_\nu(z_1)} \over {S_\nu(z_0)}} \Biggr] = \Biggl[{{1 + z_0} \over {1 + z_1}} 
\Biggr]^{-(\alpha+1)} ~~ \Biggl[{{D_L (z_0)} \over {D_L (z_1)}} \Biggr]^2, \eqno{(3)}$$

\noindent where $\rm z$ is the redshift, $\rm{S_\nu}$ is the flux density, and $\rm{D_L}$ is the
luminosity distance. In our case the reference is LLRD J1047+0739, for which
$z_0$ = 0.16828, $D_L$ = 839 Mpc and $\alpha$ = --0.85. We will derive the 
expected flux density for a source located at $\rm z$ = 5, the value at which the density of LRDs peaks.
For this $\rm z$ we have $\rm{D_L}$ = 47660 Mpc. We then obtain

$$\rm{S_\nu(z = 5) = 4.0 \times 10^{-4} ~~ S_\nu(z = 0.1682).} \eqno{(4)}$$

The expected flux densities in the centimeter range are very small, of order
50 nJy. These small flux densities will be
detectable with long integrations by the next generation
of radio interferometers.
In its band 2 (3.5-12.3 GHz)
the ngVLA (next generation VLA) will reach an rms of $\sim$14 nJy in an on-source
integration time of 100 hours (Rosero et al. 2021).

\section{Summary and Conclusions}

We present the analysis of archive VLA observations with the purpose of detecting radio continuum 
emission in LLRDs that could help understand the radio silence of cosmological LRDs.
We detect a source near J1025+1402 that could be tracing 
a compact jet emanating from an IMBH/SMBH.
We also detect radio emission from the LLRD J1047+0739
in two epochs. The nature of the emission is optically-thin synchrotron
based on the in-band spectrum.
The lack of significant time variability ($\geq$20 \%) between the two epochs (2010 and 2018)
suggests that the
radio emission can be understood in terms of a black hole of similar nature
than those found in Seyfert galaxies or by the emission produced by past
generations of supernovae.
%radio luminous supernova.
%The increase in time of the flux density suggests either two independent supernovae or
%a single supernova remnant with a radio re-brightening. 
If the radio emission from
cosmological LRDs is similar in luminosity to that produced by LLRDs
it will be possible to
detect them with long integrations with the next generation of 
centimeter interferometers. 
 
%%%%%%%%%%%%%%%%%%%%%%%%%%%%%%%%%%%%%%%%%%%%%%%%%%%%%%%%%%%%%%
\begin{acknowledgements}
 
Data calibration was performed using the NRAO VLA CASA pipeline
\footnote{https://science.nrao.edu/facilities/vla/data-processing/pipeline}.  
      LFR acknowledges support from grant CBF-2025-I-2471 of SECIHTI, Mexico.
      IFM acknowledges support from DAP-CEA-Saclay for allowing the use of software facilities and institutional access to non-open access publications.
\end{acknowledgements}

\end{document}